# Cluster-Spin Gaussian Model for Lattice- Ising Models

You-Gang Feng

Department of Basic Sciences, College of Science, Guizhou University, Cai-jia Guan, Guiyang, 550003 China
E-mail: ygfeng45@yahoo.com.cn

In this paper we lay special stress on analyzing the topological properties of the lattice systems and try to ovoid the conventional ways to calculate the critical points. Only those clusters with finite sizes can execute the self similar transformations of infinite hierarchies. Each ordered cluster has fractal dimension, their minimum relates to the edge fixed point, which accords with the transformations fixed point relating to a critical point. There are two classes of systems and clusters by their connectivity. Using mathematic mapping method we set up cluster-spin Gaussian model solved accurately. There are single state and $k$-fold coupling state in a reducible cluster, each of which corresponds to a subsystem described by a Gaussian model. By the minimal fractal dimension a final expression of the critical points is obtained. The critical points of three lattice systems are calculated, our method makes the calculations very simplified and highly accurate. A possibility of existence of different clusters at the critical temperature is discussed.

## 1. Introduction

The first exact result for the 2-dimensional Ising model was obtained by Kramers and Wannier [1]. They were followed by Onsager who derived an explicit expression for the free energy in the absence of an external field and therefore established the precise nature of the specific heat singularity [2]. A transfer matrix method was employed by these authors, and in the calculating processes the periodic boundary conditions must be used. Unfortunately, the conditions changed topological properties of the original systems as are mentioned by [3]. For example, a plane square lattice system with periodic boundary is exactly a torus square lattice system, which is not as simply connected as the original lattice. That such a system becomes ordered corresponds to a non-vanishing vector (spin) field on the torus rather than an independent spin on the plane. Moreover, it is proved that the original system itself will not automatically turn into a system with the periodic boundary in thermodynamic equilibrium [3], which means that the conditions revise the original partition function and the original free energy, so the resulting critical point should differ from the original one, although there is few difference between these datum, but it is certain to exist, which will be seen in this paper. Widom was first pointed out that, as the distance from the critical point is varied, thermodynamic functions change their scale but not their function form [4], an idea of scaling was proposed. Kadanoff applied the idea to the Ising model and in so doing opened the way for the modern theory of critical phenomena introduced by Wilson [5-7]. A concept of cluster spin was introduced by them. A cluster spin is an ordered block containing lattices and keeping the symmetries of the original system. The method introduced a concept of self similar transformations. The renormalization group theory indicates that the critical point corresponds to the fixed point of the self similar transformations. In their work it is necessary to calculate interactions of lattice spins in a cluster. The more great the size of a cluster, the more great and complex the work, which forced authors to select some approximate methods. Two rules are often used, the coarse graining and the decimation, the character is to obtain a cluster spin by making use of the summation of lattice spins in the cluster. These approximate methods resulted in that the authors never give us any highly accurate solutions. We think that for a system its physical laws relate tightly to its topological properties, if we violate its topological patterns we will not obtain real physical characters. In this paper we try to avoid those conventional ways. We lay special stress on analyzing the topological properties of the clusters. In section 2, using the fixed point theory we discuss the hierarchies for the self similar transformations. In section 3 we divide lattice systems and clusters into two classes by their connectivity. We define fractal dimension for a cluster in section 4. In section 5, giving up Wilson's supposition that the cluster spins always equal the lattice spin, we obtain relations among cluster spins, fractal dimensions and coordination numbers. In section 6 using mathematic mapping method we set up a cluster-spin Gaussian model solved accurately, and find the critical point linking to the cluster spin and its coordination number. In section 7 we analyze topologically the reducible clusters and find there are two types of spin states in a reducible cluster. In section 8 we proved the edge fixed point leads to the minimal fractal dimension, and obtained a final expression of the critical point by the minimal fractal dimensions and calculated the critical points for three lattice systems, the results indicate that our method makes the calculations very simplified and highly accurate. A possibility of existence of different clusters at the critical temperature is discussed in section 9, and conclusion in section 10.

## 2. Hierarchies of the self similar transformations

We are concerned about what type of transformations will have a fixed point. Let the clusters sizes and shapes to be identical, so the system can make the self similar transformations entirely. Let the distance of interaction between two nearest neighbor cluster spins $A$ and $B$ be $d(A,B)$, after a self similar transformation they become nearest neighbor lattice spins $f(A)$ and $f(B)$, and the distance of interaction between neighbors be $d(f(A), f(B))$, the Lifschetz constant $L$ be defined by [8]

$$L = \frac{d(f(A), f(B))}{d(A, B)} \tag{1}$$

If the cluster size is limited, i.e. $d(A,B) < \infty$, so that $0 < L < 1$ for $d(A,B) > d(f(A), f(B))$. If the cluster size approaches infinity, so $d(A,B) \to +\infty$ and the constant tends to zero, $L \to 0$. According to Lifschetz fixed point theorem [8,9], there is not any fixed point in the self similar transformations for $L \to 0$. If and only if the cluster size is limited the constant $L$ will certainly be smaller than one and not vanish, $0 < L < 1$, so that the self similar transformations of the cluster spins have a unique fixed point. At that time the correlation length changes into infinity only by infinite iterations, which shows that the self similar transformations are under the necessity of hierarchies, and on the infinite hierarchy a system will become an ordered cluster. We call an original lattice the zeroth order lattice, they construct a first order cluster by the transformation, and the cluster is said to be on the first hierarchy of the transformations. On the $m$ th hierarchy there must be only the $m$ th order clusters which are independent of each other, and a $m$ th order cluster contains the ($m$-1)th order lattices which are just the ($m$-1)th order clusters before rescaling. It is clear that the inside space of the $m$th order cluster is just the outside space of the ($m$-1)th lattices, whish is of dimensions $D$ (see section 4). As a lattice, the inside of a ($m$-1)th order lattice is indistinguishable.

### 3. Two classes of systems and clusters

There are varieties of lattice systems, their clusters differ from each other, what characters the clusters will have? A carrier space of an ordered cluster must be simply connected tantamount to a single point space, because of which an ordered cluster can shrink to a lattice. This character is a mathematic basis for the rescaling cluster spins. Topology point out that only triangles and tetrahedroids are simply connected, they are mathematically called simplexes, without relating to their sizes [10]. By connectivity we divide lattice systems into two classes: The first, an irreducible system to which only the plane triangle lattice and the tetrahedroid lattice belong. The second, a reducible system, to which all of lattice systems belong apart from the first. In the irreducible system a cluster which keeps the system symmetries is called irreducible cluster, the triangle clusters and the tetrahedroid clusters are of them. Some identical irreducible clusters can directly form a new larger irreducible cluster, making the transformations go on. In the reducible system a cluster which keeps the system symmetries is called reducible cluster, of which a square lattice and a cube lattice are. Although a reducible cluster keeps all of the symmetries of the original system, but its carrier space is a complex in the mathematic sense, which is not simply connected. Mathematically, a complex can be decomposed to some simplexes, each of which is simply connected and is called a subcluster of the reducible cluster. The subclusters shape must differ from the reducible cluster one, otherwise they are merely the reducible clusters with different sizes and the same topological properties [10]. The subclusters keep only partial symmetries of the original system, if they execute directly the transformation the original symmetries will be broken, thus such a transformation is impossible. The self similar transformations of a reducible system take two steps: First, $k$ subclusters formed in a reducible cluster. Second, the $k$ subclusters make the reducible cluster ordered by their interactions. In fact, the two steps occur simultaneously. After that, some ordered reducible clusters can construct a new bigger reducible cluster, making the self similar transformations on a higher hierarchy. Topology indicates that a carrier space of an ordered reducible cluster is equivalent to a product space of its subclustres' spaces [10]. A geometric grid is a carrier of lattice spins and the topological properties of the lattice spin system are attached to its carrier topology [3]. See figure 1, a triangle with vertices of $P$ has $n^2$ cells, $P = (n+1)(n+2)/2$, where a cell is a minimal simplex. If we put a spin on each vertex, the simplex then becomes a cluster spin containing $P$ lattice spins with edge $n$, where a distance of two nearest neighbors is defined as a unit length.

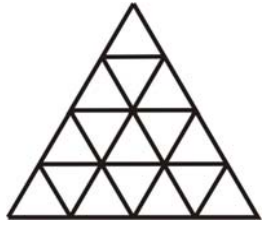

Fig. 1.   An irreducible cluster of the triangle lattice with edge $n = 4$.

### 4. Fractal dimensions

The self similar transformations have fractal properties, which leads to fractal dimensions [11]. Wilson did not notice that a cluster has a fractal structure, the difficulties in calculations forced him to take some approximate methods. We think that understanding the fractals will help us research deeply the properties of the critical points. In the triangle lattice system the spin directions on the lattice sites differ from each other, such a non-uniformity of the orientations makes a lattice different from others, thus a cluster has a fractal dimension. Let a cluster with edge $n$ be covered by open balls with diameter $1/n$, which number be $P$ at least, then the fractal dimension $D$ of the cluster is defined as [11,12]

$$D = -\frac{LnP}{Ln(1/n)} = \frac{LnP}{Ln(n)} \qquad (2)$$

Both a cluster and a fractal dimension are produced in the self similar transformation, as mentioned in section 1 the cluster is an ordered block, so only the cluster has the fractal dimension. A disordered block acts as an empty set in the transformation process, and there is no relation between a disordered block and a fractal dimension. The values of the edges should guarantee the fractal dimension meaningful, otherwise (2) has no meaning. By the definition (2), the inside space of a cluster amounts to a super cube of dimensions $D$ with edge $n$ and volume $P = n^D$, in this case the fractal dimensions are also called capacity dimensions [11,12]. By (2), a fractal dimension $D_{tr}$ for the irreducible cluster of the triangle lattice is

$$D_{tr} = \frac{Ln[(n+1)(n+2)/2]}{Ln(n)} \qquad (3)$$

From (3) we see that $D_{tr}$ is an edge function. For the plane square lattice, since there is no next-nearest neighbor interaction in Ising model the carrier of a reducible cluster cannot be decomposed to some triangles, as usual. However, as the carrier is complex, it will be decomposed certainly to some special shapes. Let there be $k$ subclusters in a reducible cluster, from (2) a fractal dimension $D_{sq}$ of a subcluster of the square lattice be defined by

$$D_{sq} = \frac{Ln[(n+1)^2/2]}{Ln(n)} \quad , \qquad (4)$$

where $k = 2$. In figure 2, $k = 3$, the subcluster intervening between two subclusters cannot transform in the same way as the other subclusters do. When the edges of its neighbor subclusters change into half-infinity, the edge of the middle subcluster has to keep limit, such non-uniform transformation will break the original symmetries. If $k = 4$, there are four squares with smaller edges which are complexes equally. The cases of $k > 4$ are similar to the cases of $k = 3$ or $k = 4$. Similarly, the fractal dimension $D_{cu}$ of a subcluster of the cube lattice is

$$D_{cu} = \frac{Ln[(n+1)^3/4]}{Ln(n)} \tag{5}$$

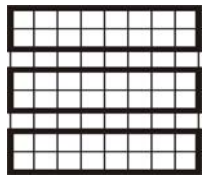

Fig.2. A reducible cluster containing three subclusters of the square lattice with $n = 8$.

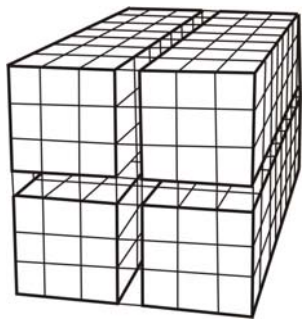

Fig. 3. A reducible cluster with four subclusters of the cube lattice with $n = 7$.

As illustrated in figure 3, a cube is subdivided into four cuboids, $k = 4$. According to simplicial decomposition theory a three dimensional complex should be composed of three dimensional simplexes, namely, it must be decomposed in three dimensional directions, so $k \neq 2$. If $k = 8$ there are 8 cubes with smaller edges which are complexes still. If $k = 3$, 5, 6, 7 or $k > 8$ the transformations will break the system symmetries. Thus the case of $k = 4$ is correct. The changing of the fractal dimension reflects directly the self similar transformations. Since the fractal dimension is an edge function the edges fixed point certainly accords with the transformations' one.

## 5. A relation of the cluster spin with its fractal dimension

Now that different sizes accord with different fractal dimensions, for those clusters with different fractal dimensions what values of spins they will have? Let a cluster spin be $S$, the energy of interaction between two nearest neighbors denoted by $y_1$ and $y_2$ be $JS^2$, where $J$ be a coupling constant; as the self similar transformation a new cluster formed on higher hierarchy the two cluster spins become two nearest neighbor lattice spins in the new cluster after rescaling, and denoted by $f(y_1)$ and $f(y_2)$ with each magnitude of $s$, $s^2 = 1$, and their interacting energy be $js^2$, where $j$ be a coupling constant in the $D$-dimensional space, the inside space of the new cluster. Let $d(f(y_1), f(y_2)) = js^2$ and $d(y_1, y_2) = JS^2$. The self similar transformation is virtually a contraction mapping [9], we then have a equation similar to (1)

$$d(f(y_1), f(y_2)) = r \cdot d(y_1, y_2) \quad, \tag{6}$$

$$js^2 = rJS^2 \tag{7}$$

(7) shows that there is a quantitative relationship between $js^2$ and $JS^2$. Note that

the outside space of a cluster is its embedded space, which always is the Euclidean space of dimensions $N$. After the self similar transformations and rescaling on the $m$ th hierarchy a cluster which was originally called the ($m$-1)th order cluster has changed into a new lattice, which is called a ($m$-1)th order lattice. As a lattice it lies the inside of the $m$ th order cluster, so its outside space is just the inside space of the $m$ th order cluster of dimensions $D$. For an observed object, whenever it serves as a lattice spin the interacting energy equals $js^2$; as a cluster spin, however, the energy is $JS^2$. As mentioned in section 4.1, a lattice in a cluster of dimensions $D$ can be equivalently regarded as a lattice in a super cube of dimensions $D$ with edge $n$. It is well known that a coordination number for a $D$-dimensional cube is $2D$, so that the total magnitudes of interacting energies of a lattice spin with all its nearest neighbors inside the $m$ th order cluster equal $2Djs^2$. As a ($m$-1)th order cluster spin before rescaling, however, the total interacting energies of it with all its nearest neighbors, in its outside space, equal $ZJS^2$, where $Z$ is a coordination number of the cluster spin. In fact, the lattice spin and the cluster spin are the same observed object described by two above artificial versions, hence these different descriptions must be equivalent in magnitudes, which leads to an equality

$$ZJS^2 = 2Djs^2 \quad , \tag{8}$$

where $s^2 = 1$. Comparing (8) with (7), we find

$$r = Z/(2D) \tag{9}$$

As the constant $L$ in (1), $r$ is limited indeed. Let $K = J/(k_B T)$, $k_B$ the Boltzman constant, $T$ temperature, (8) becomes

$$ZKS^2 = 2Djs^2/(k_B T) \tag{10}$$

For a given system, $Z$ is constant, $s^2 = 1$, $j$ is the coupling constant in the space of dimensions $D$ relating with $D$, which means $K$ and $S^2$ also relate to $D$, and they are not independent of each other. In renormalization group theory [6], Wilson preferred to suppose $S^2 = s^2$, namely, the cluster spins always will not change, so that $K$ is only determined by $D$ at the critical temperature due to (10). Since the fractal dimension $D$ is an edge function, $K$ is certainly controlled by the edges. In fact, $K_c$ is merely determined by the edge fixed point, which will be proved in section 8. We think that there is no enough evidence to guarantee the supposition of $S^2 = s^2$ to be correct for any systems and clusters no matter what size and shape the cluster will have, and what type of systems such as the plane triangle lattice, the plane square lattice or the cube lattice the cluster lies in. The cluster spins' coupling is virtually some lattice spins' one, which are shown in (7) and (8). Thus we have sufficient reasons to suppose that $J = j$ and $S^2 \neq s^2 = 1$, so (8) becomes

$$ZS^2 = 2D \qquad (11)$$

(11) indicates that different cluster spins with $D$ and $Z$ have different values. We emphasize here that we derived (8), (10) and (11) by the same observed object, so the $J$ and $j$ in (8) are only fit for the same system. Each subsystem of a reducible system has its own equalities similar to (8) and (11). The suitability of (10) and (11) will be discussed further in section 7.2. The supposition that the cluster spins can change gives us a chance to set up a cluster-spin Gaussian model solved exactly, which will make the calculation of a critical point very simplified and highly accurate.

**6. Cluster-spin Gaussian model**
**6.1. Partition function and free energy**
On the one hand all of clusters with finite edges can exert infinitely the transformations, on the other hand the fixed point is unique, so we infer that there is a special edge which accords with the fixed point. The special edge determines a special fractal dimension, which will be proved in the following Gaussian model. When we approach a critical point the correlation range becomes larger and larger, but always limited, which leads to a finite hierarchy of the self similar transformations, thus we may consider the cluster spins independent variables provided that the correlation range is not larger than the size of a cluster. According to Ergodic hypothesis there are clusters with a variety of shapes and sizes in thermodynamic equilibrium, so there are two cases in the original system: The first case, the system contains infinite identical ordered clusters and they keep the original symmetry, apart from them, there are either disordered regions or empty sets. In figure 5 three triangles are the nearest neighbor ordered clusters, among them the blank space represents a disordered region having lattice spins less than an ordered cluster has. In order to keep the symmetries and make the self similar transformations to higher hierarchy the system forbid two ordered clusters with different sizes to be at the same time. There is no interaction of a disordered region and an ordered cluster, and so we view each disordered region as an empty set. In this case the system can make self similar transformations infinitely. The second, the cases apart from the first, there are no infinite self similar transformations, in the view of which we consider only the first case. Now we set up a new system with infinite lattices, the orientations of the new lattice spins are up or down at random, the magnitudes of all the new lattice spins are equal and can change. The new system itself cannot execute any transformations. First, the new system keeps the symmetry of the original system. Second, as the ordered clusters' sizes are either large or small, no matter what sizes the cluster will have a new lattice always represents an ordered cluster. This fact means that the values of a new lattice spin involve the values of all possible cluster spins. Last but not least, no matter what size the ordered cluster will have, we always rescale the distance between two nearest neighbors a constant, which is just the distance of adjacent new lattice spins. We notice that the distances mentioned above do not contribute to the critical point, thus we can determine the constant arbitrarily. It is obvious that the new system has the same critical point as lattice- Ising model. It is also clear when we discuss the self similar transformations we analyze lattice-Ising model, when we study the critical point we use the new system. The following calculation is similar to [13-15], when we mention "lattice" or "lattice spin", we mean the lattice or the lattice spin of the new system.

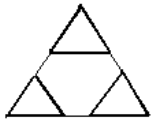

Fig.5. A disordered region is among three ordered clusters.

On the $m$ th hierarchy the orientations of the cluster spins are different from each other, namely, the correlation length is less than the cluster size, each cluster spin is viewed as an independent variable, so we can use the statistical laws to describe the system. The space of interaction of two adjacent lattice spins is the Euclidean space of dimensions $N$. In the cluster-spin Gaussian model of the plane triangle lattice, $N = 2$, the dimensions of their lattice vector space (Bravais lattice space) and reciprocal lattice vector space (wave vector space) also are $N = 2$, respectively. The methods of calculating bases of lattice vector space, bases of reciprocal lattice vector space and Brillouin zones are commonly applied in solid state physics [16,17]. A solid state physical cell is drawn in figure 6. The bases are given by

$$\vec{a_1} = a\vec{i}, \qquad \vec{a_2} = (a/2)[\vec{i} + \sqrt{3}\vec{j}], \qquad (12)$$

where a cell edge is denoted by $a$, which also is designated as a lattice constant, $\vec{i}$

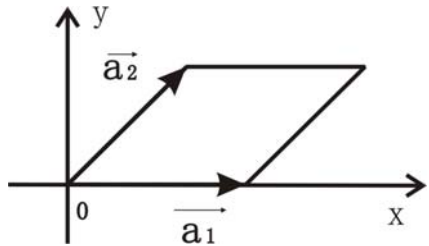

Fig.6.   A solid state physical cell of the cluster-spin
Gaussian model of the plane triangle lattice.

and $\vec{j}$ are unit vectors for $x$-axis and $y$-axis, respectively. The bases of the reciprocal lattice vector space are given by

$$\vec{b_1} = (2\pi/a)[\vec{i} - (1/\sqrt{3})\vec{j}], \qquad \vec{b_2} = (2\pi/a)(2/\sqrt{3})\vec{j} \qquad (13)$$

The number of the reciprocal lattices with the same distance $(4\pi/a)(1/\sqrt{3})$ from the origin equals 6, which is just the coordination number, and their coordinates are $[\pm 2\pi/a, \mp 2\pi/(\sqrt{3}a)]$, $[0, \pm 4\pi/(\sqrt{3}a)]$ and $[\pm 2\pi/a, \pm 2\pi/(\sqrt{3}a)]$. The first Brillouin zone is illustrated as a hexagon in figure 7. The region over which the values of the bases $q_x$ and $q_y$ of the reciprocal lattice vector space go over from the negative to the positive are expressed as

$$-4\pi/(3a) \leq q_x < 4\pi/(3a), \qquad -2\pi/(\sqrt{3}a) \leq q_y < 2\pi/(\sqrt{3}a) \qquad (14)$$

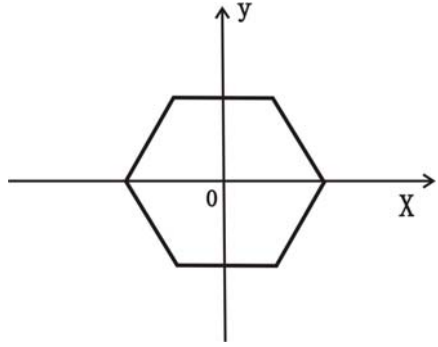

Fig.7. The first Brillouin zone of the cluster-spin Gaussian model of the plane triangle lattice.

In the absence of an external field the Hamiltonian of the system takes

$$H = K\sum_{(i,j)} S_i S_j, \qquad (S_i, S_j = \pm S_{tr}), \qquad (15)$$

where $K \equiv J_{tr}/(k_B T)$, $J_{tr} = J = j$, $S_{tr}$ is determined by (10), $\sum_{(i,j)}$ is sum over all possible nearest neighbor configurations. Without any consideration of fluctuations, we get the partition function

$$Q^* = \sum_{(i,j)} \exp H \qquad (16)$$

From (3) and (11), we know that $S_i$ and $S_j$ are determined by the clusters edges, which fluctuations should obey the central limit theorem in thermodynamic equilibrium [15,18], so the fluctuations of the spins can be described by Gaussian distribution

$$W = \prod_{j=1}^{N_c} \exp[-S_j^2/(2\langle S_{tr}^2 \rangle)] \ , \tag{17}$$

where $N_c$ is the total number of cells, $N_c \to +\infty$, and $\langle S_{tr}^2 \rangle$ is the mean square of the lattice spins. For simplicity, we extend the range of the magnitudes of $S_{tr}$ to infinity, $-\infty < S_{tr} < +\infty$, and make the values continuous. Such a procedure is based on that we concentrate only on a singularity of the free energy determined by the partition function rather than on the function values. The later computing results will reveal that the meaningful values of $S_{tr}$ are absolutely controlled by (3) and (11), independent of the range extension. The Gaussian weighting factor gives the largest weight to the cluster spin value $S_j = 0$, which means that those clusters have inadmissible edges and the clusters have not any fractal dimension (for example, in the plane triangle lattice system when $n = 1$ ). Considering (15)-(17), we see that the partition function is of the form

$$Q = \int_{-\infty}^{+\infty} \cdots \int_{-\infty}^{+\infty} \prod_{j=1}^{N_c} dS_j \exp\{K \sum_{(i,j)} S_i S_j - \frac{1}{2\langle S_{tr}^2 \rangle} \sum_{j=1}^{N_c} S_j^2\} \tag{18}$$

The integration is only carried out in the inscribed circle of the first Brillouin zone. As mentioned by [14,15], and seen follows, such a calculation is still valid because a contribution to the singularity of the free energy uniquely comes from the range near the origin covered by the inscribed circle, independent of the magnitudes of the lattice vectors. In other words, the singularity depends only on the long-wavelength part of the wave vectors. In follows, we introduce Fourier transformations of the spins,

$$S_i = \frac{1}{\Omega} \sum_q S_q \exp(i\vec{q}\cdot\vec{r_i}) \ , \qquad S_q = V \sum_{i=1}^{N_c} S_i \exp(-i\vec{q}\cdot\vec{r_i}) \ , \tag{19}$$

where $V$ is a cell volume, $\Omega = N_c V$ is the total volume of the system, $N_c$ the total number of cells and the sum over $\vec{q}$ is restricted in the first Brillouin zone. Let

$$K(\vec{r_i}-\vec{r_j}) \equiv \begin{array}{l} K, \quad \vec{r_i},\vec{r_j} \ nearest \\ 0, \quad \quad others \end{array} \tag{20}$$

and

$$H = \frac{1}{2}\{\sum_{i,j} K(\vec{r_i}-\vec{r_j})S_i S_j - \frac{1}{\langle S_{tr}^2 \rangle} \sum_{i,j} \delta(\vec{r_i}-\vec{r_j})S_i S_j\} \ , \tag{21}$$

where the two terms $\sum_{i,j}$ are, different from $\sum_{(i,j)}$ , independent sums over $i$

and $j$, respectively, and

$$\delta(\vec{r_i}-\vec{r_j}) \equiv \begin{matrix} 1, & \vec{r_i}=\vec{r_j} \\ 0, & others \end{matrix} \qquad (22)$$

Consider a Fourier transformation of the form as

$$\sum_{i,j} f(\vec{r_i}-\vec{r_j})S_i S_j \qquad (23)$$

Using (19), we make (23) be

$$\sum_{i,j} f(\vec{r_i}-\vec{r_j})S_i S_j = \sum_{i,j} f(\vec{r_i}-\vec{r_j})\frac{1}{\Omega}\sum_{q_1} S_{q_1} \exp(i\vec{q_1}\cdot\vec{r_i})\cdot\frac{1}{\Omega}\sum_{q_2} S_{q_2} \exp(i\vec{q_2}\cdot\vec{r_j})$$

$$= \frac{1}{\Omega V}\sum_{q_1,q_2} S_{q_1} S_{q_2} \frac{1}{N_c}\sum_i \exp[i(\vec{q_1}+\vec{q_2})\cdot\vec{r_i}]\sum_j f(\vec{r_i}-\vec{r_j})\exp[-i\vec{q_2}\cdot(\vec{r_i}-\vec{r_j})] \qquad (24)$$

Let

$$f(\vec{q_2}) = \sum_j f(\vec{r_i}-\vec{r_j})\exp[-i\vec{q_2}\cdot(\vec{r_i}-\vec{r_j})] \quad , \qquad (25)$$

where $f(\vec{q})$ and a Fourier component of $f(\vec{r})$ are a constant apart at most. Noticing an orthonormality

$$\frac{1}{N_c}\sum_i \exp[i(\vec{q_1}+\vec{q_2})\cdot\vec{r_i}] = \delta(\vec{q_1}+\vec{q_2}) \equiv \begin{matrix} 1, & \vec{q_1}+\vec{q_2}=0 \\ 0, & others \end{matrix} \quad , \qquad (26)$$

where $\vec{q_1}$ and $\vec{q_2}$ take all possible values and are renewedly $\vec{q_2}=\vec{q}$, $\vec{q_1}=-\vec{q}$. $S_i$ are real numbers, $S_q^*=S_{-q}$, where $S_q^*$ is conjugate to $S_q$. (24) becomes

$$\sum_{i,j} f(\vec{r_i}-\vec{r_j})S_i S_j = \frac{1}{\Omega V}\sum_{q_1}\sum_{q_2} S_{q_1} S_{q_2} f(\vec{q_2})\delta(\vec{q_1}+\vec{q_2}) = \frac{1}{\Omega V}\sum_q f(\vec{q})|S_q|^2 \qquad (27)$$

Using (27), analogously, we get

$$\sum_{i,j} K(\vec{r_i}-\vec{r_j})S_i S_j = \frac{1}{\Omega V}\sum_q K(\vec{q})|S_q|^2 \qquad (28.1)$$

From (25) and (20), similarly, we have

$$K(\vec{q}) = \sum_j K(\vec{r_i}-\vec{r_j})\exp[-i\vec{q}\cdot(\vec{r_i}-\vec{r_j})] = K\sum_{\vec{\delta}_{ij}} \exp(-i\vec{q}\cdot\vec{\delta}_{ij}), \qquad (28.2)$$

where $\vec{\delta}_{ij}$ is a vector from the lattice $i$ to its nearest neighbor lattice $j$, (28.2) is ready to calculate the critical point, see (32). Using (27) and inserting (28.1) to (15), noticing that whenever $f(\vec{r}_i - \vec{r}_j) = \delta(\vec{r}_i - \vec{r}_j)$, $f(\vec{q}) = 1$, we have (15) take the form

$$H = \frac{1}{2\Omega V}\sum_q [K(\vec{q}) - \frac{1}{\langle S_{tr}^2 \rangle}]|S_q|^2 \tag{29}$$

(29) is inserted to (18), a typical Gaussian integration is obtained

$$Q = \int_{-\infty}^{+\infty}\cdots\int_{-\infty}^{+\infty}\prod_q dS_q \exp\{-\frac{1}{2\Omega V}\sum_q [\frac{1}{\langle S_{tr}^2 \rangle} - K(\vec{q})]|S_q|^2\} = \prod_q \left[\frac{2\pi\Omega V}{\frac{1}{\langle S_{tr}^2 \rangle} - K(\vec{q})}\right]^{1/2} \tag{30}$$

Thus, the free energy determined by $Q$ is of the form

$$F = -k_B T LnQ = \frac{1}{2}k_B T \sum_q Ln[\frac{1}{\langle S_{tr}^2 \rangle} - K(\vec{q})] + T \cdot const. \tag{31}$$

**6.2. Determination of the critical point**

From (31), the singularity of the free energy turns up when $K(\vec{q}) = 1/\langle S_{tr}^2 \rangle$, which corresponds to the critical point. Whenever the temperature $T$ is higher than $T_c$, $K(\vec{q})$ always is less than that of $1/\langle S_{tr}^2 \rangle$, so the approach, $K(\vec{q}) \to 1/\langle S_{tr}^2 \rangle$, results in the maximum of $K(\vec{q})$ at the critical temperature, inversely the minimum of $\langle S_{tr}^2 \rangle$, $\langle S_{tr}^2 \rangle_{min}$. It should be emphasized that $\langle S_{tr}^2 \rangle_{min}$ is directly a solution of the singularity of the free energy and is regarded as a result of the mutation of statistical regular pattern rather than a statistical mean value at the critical temperature. For the triangle lattice the coordinates of six lattice vectors associated with $\vec{\delta}_{ij}$ in (28.2) around the origin are given by $(\pm a, 0)$, $(\pm a/2, a\sqrt{3}/2)$, and $(\pm a/2, -a\sqrt{3}/2)$. Inserting these to (28.2) yields

$$K(\vec{q}) = 2K[\cos(q_x a) + \cos(q_x a/2 + q_y a\sqrt{3}/2) + \cos(q_x a/2 - q_y a\sqrt{3}/2)] \tag{32}$$

$K(\vec{q})$ reaches its maximum at $\vec{q} = 0$, thus we immediately get

$$K(0) = 6K_c = 1/\langle S_{tr}^2 \rangle_{min} \tag{33}$$

The whole process of deriving the critical point formula reveals that the reciprocal lattice vectors always vanish at the critical point. From (32), it is clear that the number 6 in (33) is just the coordinate number of a cluster spin. Thus, for any system whenever we know its coordination number $Z$ and the minimal average square value $\langle S^2 \rangle_{min}$, we can get its critical point expression

$$K_c = \frac{1}{Z <S^2>_{min}} \quad , \tag{34}$$

where $<S^2>_{min}$ will be determined by a special edge.

## 7. Two types of spin states in a reducible cluster
### 7.1. Coupling states and critical points

A cluster spin is a result of its subcluster spins interactions, and so the state of the cluster spin differs from its subcluster spin stae. We call the state of a subcluster spin a singlet state, there are $k$ single states. The $k$ subcluster spins interact on each other making the reducible cluster ordered, which spin state is called a $k$-fold coupling state. For the first state, each subcluster spin preserves its own independence and equals a single point space resulted from their simply connectivity, at the moment the reducible cluster is disordered still and cannot be contractible. The correlation length is not greater than a range the size of a subcluster, and so the outside space of a subcluster is still the Euclidean space of dimensions $N$. For example, in the plane square lattice system, $k=2$, there are two subclusters, each of which is in the singlet state. As a singlet state a subcluster spin is $S_{11}$, its coordination number is $Z_{11}=2$.

For the second state, the reducible cluster becomes ordered, for the moment it acts as a simply connected space and the correlation range is as large as its size. According to topology [10], only a product space of $k$ connected spaces is connected, which dimensions equal $D^k$, where $D$ is the dimension of a subcluster space. Of course, such a version is only a topologically equivalent description of an ordered reducible cluster. Whenever a reducible cluster becomes ordered it will be able to shrink, so its coordination number is just the original one of the system. For example, in the plane square lattice system, a reducible cluster spin is $S_{12}$, its coordination number $Z_{12}=4$.

Thus, on a hierarchy there exist simultaneously two independent types of spin states, which should correspond to two statistically independent subsystems, each of which can be described by a Gaussian model. Let the partition function of the first independent subsystem be $Q_{11}$, in which a lattice spin be $S_{11}$ with coordination number $Z_{11}$; the partition function of the second independent subsystem be $Q_{12}$, a lattice spin $S_{12}$ with coordination number $Z_{12}$. The partition function $Q_{sq}$ of the plane square lattice system is given by

$$Q_{sq} = Q_{12} Q_{11} \quad , \tag{35.1}$$

where the product form of $Q_{11}$ and $Q_{12}$ implies that there exist two statistically independent events simultaneously. A discussion analogous to the irreducible system in sections 6.1 and 6.2 indicates that the critical point of the system depends only on $Q_{11}$ and $Q_{12}$. We get a logarithmic form of the partition functions associated with the free energy:

$$LnQ_{sq} = LnQ_{11} + LnQ_{12} \tag{35.2}$$

By (31) and (35.2), we see that the singularity of the free energy is determined by both of the singularities of $LnQ_{11}$ and $LnQ_{12}$. Let the critical point of the system be $K_c$ relating to the singularity of $LnQ_{sq}$, the critical point $K_{c1}$ to the singularity of $LnQ_{11}$, the critical point $K_{c2}$ to the singularity of $LnQ_{12}$. (35.2) shows us that $K_c$ should be the sum of $K_{c1}$ and $K_{c2}$, so

$$K_c = K_{c1} + K_{c2} \tag{36}$$

Thus we can obtain the critical point of a reducible system such as the plane square lattice system by computing the critical points of its subsystems.

### 7.2. Quantitative calculations of reducible cluster spins

Both of the $k$ singlet states and one $k$-fold coupling state lie on the same hierarchy. Thus, a $m$ th order subcluster on the $m$ th hierarchy should shrink to a lattice in a $(m+1)$ order subcluster on the $(m+1)$th hierarchy. For example, in the first independent subsystem of the plane square lattice system ($k = 2$), a $m$ th order subcluster spin $S_{11}$ (the singlet state) with coordination number $Z_{11}$ and coupling constant $J_{11}$ becomes a lattice spin $s$ with coupling constant $j_{11}$ in a space of dimensions $D_{sq}$ after rescaling, see (7), the space is the inside space of the $(m+1)$ th order subcluster which is equivalent to a super cube of edge $n$, volume $n^{D_{sq}}$ and coordination number $2D_{sq}$. By (11), we get

$$Z_{11}S_{11}^2 = 2D_{sq} \tag{37.1}$$

The same is true for the second independent subsystem, we find

$$Z_{12}S_{12}^2 = 2D_{sq}^2 \;, \tag{37.2}$$

where a $m$ th order reducible cluster spin $S_{12}$ with coordination number $Z_{12}$ and coupling constant $J_{12}$ becomes a lattice spin $s$ with coupling constant $j_{12}$ in a $(m+1)$ th order reducible cluster after rescaling, the inside space of the $(m+1)$ th order reducible cluster is equivalent to a super cube of dimensions $D_{sq}^2$ with coordination number $2D_{sq}^2$. (37.1) and (37.2) imply our suppositions of $J_{11} = j_{11}$ and $J_{12} = j_{12}$. However, we can not infer that $J_{11} = j_{12}$, $J_{12} = j_{11}$, or $J_{11} = J_{12}$, since these constants belong to two different subsystems, the constants of different subsystems are incomparable with each other. With the same reason, in the cube lattice system, for its first independent subsystem we get

$$Z_{21} S_{21}^2 = 2D_{cu} \quad , \qquad (38.1)$$

where $S_{21}$ represents a subcluster spin (the singlet state) with coupling constant $J_{21}$ and coordination number $Z_{21}$, $D_{cu}$ determined by (5), and further

$$Z_{22} S_{22}^2 = 2D_{cu}^4 \quad , \qquad (38.2)$$

where $S_{22}$ denotes a reducible cluster spin (fourfold coupling state) with coupling constant $J_{22}$ and coordination number $Z_{22}$. Its inside space is equivalent to a super cube of dimensions $D_{cu}^4$, which is a product space of its four subcluster spaces, each of which is of dimensions $D_{cu}$. $J_{21}$ and $J_{22}$ are incomparable as they belong to different subsystems. In the cube system, there are no $k$-fold coupling states for $k=2$, or 3 resulted from two or three subcluster spins interactions. If the states were reasonable, the additional subsystems would increase the system free energy.

## 8. Numerical calculations of the critical points by $D_{\min}$

Since the spin $S$ in the $\langle S^2 \rangle$ must be in the range over which the cluster spins take, we then have $\langle S^2 \rangle = S_c^2$, where $S_c$ is determined by (11). Because $\langle S^2 \rangle$ should be the minimum at the critical point, $S_c$ depends only on the minimum of the fractal dimensions $D_{\min}$. For (3), the zero value for the derivative of $D_{tr}$ with respec to $n$, $dD_{tr}/dn = 0$, leads to a fixed point equation for edges

$$f(n) = [(n+1)(n+2)/(2n+3)] \cdot \frac{Ln[(n+1)(n+2)/2]}{Ln(n)} = n \quad (39)$$

The equation (39) has a unique edge fixed point $n^*$ due to the Banach fixed point theorem [9,19]. Computing (3) and (39) yields the minimum

$$D_{tr,\min} = 1.814055098, \qquad n^* = 14.4955 \quad (40)$$

With the same reason, computing (4), we get a unique minimum $D_{sq,\min}$ with an edge fixed point $n^*$ for the subcluster of the plane square lattice

$$D_{sq,\min} = 1.779990992, \qquad n^* = 7.839995 \quad (41)$$

Similarly, for the subcluster of the cube a unique minimum $D_{cu,\min}$ with an edge fixed point $n^*$ is

$$D_{cu,\min} = 2.478143004, \qquad n^* = 4.749100 \quad (42)$$

Combining (11) with (34), we find a final expression of the critical points with $D_{\min}$

$$K_c = \frac{1}{2D_{\min}} + \frac{1}{2D_{\min}^k}, \quad (43)$$

where $k$ relates to a $k$-fold coupling state, the first term relates to the single state. Obviously, for an irreducible system the second term will vanish. By using (43), we can numerically calculate the critical points. For the plane triangle lattice, from (40) and (43) we get $K_c = 0.2756$. With the help of periodic boundary conditions, Kramers and Wannier got a result [1], 0.2747. For the plane square lattice, from (41) and (43) we get $K_c = 0.4387$. Making use of periodic boundary conditions, Onsager obtained a result [2], 0.4407. For the cube lattice, by (42) and (43) we have 0.2150, $u_c = th(K_c) = 0.2118$ is its hyperbolical function. [20] got an approximate result $K_c = 0.2217$, other approximate results in the forms of hyperbolical functions are given by [21], [22], and [23], they are 0.2108, 0.21813, 0.21811, respectively.

## 9. A possibility of existence of different clusters at $T_c$

Although the supposition of that there are identical clusters at the critical point helps us get highly accurate results, it doesn't eliminate a possibility of existence of different clusters at the critical temperature. The actual transformations permit only integer edges, the fractional edges $n^*$ associate only with the fixed points. From (3), (4) and (5) we know that around $n^*$ those obviously different edges, $n > n^*$ or $n < n^*$, correspond to the almost same fractals. In order to approach the critical point further a system is forced to adjust continuously its clusters' edges, which will cause great fluctuations of the edges around the $n^*$. In the adjustment process different clusters will occur.

**10. Conclusion**

Only those clusters with finite sizes can execute the self similar transformations with infinite hierarchies, which have a unique fixed point. There are two classes of clusters: irreducible clusters and reducible clusters. Each ordered cluster has fractal dimensions, which minimum corresponds to the transformations' fixed point, relating to a critical point. There are two types of spin states in a reducible cluster: single state and $k$-fold coupling state. Each type corresponds to a subsystem described by a Gaussian model. By the minimal fractal dimension a final expression of the critical points is obtained.

The fluctuations of the clusters' edges allow there to be different clusters at $T_c$. We calculated the critical points for three lattice systems: plane triangle, plane square, cube. Making use of our method the numerical calculations of the critical points become very simplified and highly accurate.